# Design and control analysis of a deployable clustered hyperbolic paraboloid cable net

Shuo Ma[a,b], Kai Lu[a], Muhao Chen[c,*], Robert E. Skelton[c]

[a] *College of Civil Engineering, Zhejiang University of Technology, Hangzhou, Zhejiang 310023, China*

[b] *Key Laboratory of Space Structures of Zhejiang Province, Hangzhou, Zhejiang 310058, China*

[c] *Department of Aerospace Engineering, Texas A & M University, College Station, Texas 77840, United States*

**Abstract:** This paper presents an analytical and experimental design and deployment control analysis of a hyperbolic paraboloid cable net based on clustering actuation strategies. First, the dynamics and statics for clustered tensegrity structures (CTS) are given. Then, we propose the topology design of the deployable hyperbolic paraboloid cable net. The deployability of the cable net is achieved by using clustered cables. It is shown that the clustered cables significantly reduce the number of actuators required for control. The deployment trajectory and actuation prestress in the cables are designed to ensure the tensions are feasible during the deployment process. Then, we compare the deployment analysis's open-loop and closed-loop control strategies. Finally, a lab-scale model is constructed to validate the actuation laws. We test the static performance and deployment process of the experimental model. Results show that the closed-loop control approach is more stable and smoother than the open-loop one in the deployment process. The approaches developed in this paper can also be used for various deployable tensegrity structures.

## 1  Introduction

For many civil structures, apart from the functionalities from the view of mechanics and structural engineering, architects always want to promote the aesthetic value of structures as best as they can. However, the balance of architect aesthetics, loading efficiency, construction practice, and budget limitations normally brings a lot of conflicts between architects and engineers. The communications between the two professionals turn out to be a hardship in the negotiation and delay in the delivery date. Both the architecture and engineering communities are looking for a better structure paradigm and more substantial materials for the next-generation engineering structures. After decades of study, tensegrity has shown its strong candidacy and may provide a solution to design efficient structures while preserving the beauty due to its many excellent properties in high stiffness to mass ratio [1], multiple self-stress states [2,3], extensive morphing capabilities [4], and tunable structural parameters [5], easy integration structure and control [6], etc. Nowadays, many studies have shown the advantages and feasibilities of building advanced structures, such as deep space habitats [7], cable domes [8], mining rigs [9], tensegrity robots [10], tensegrity fish [11], and tensegrity spines [12], etc. Recent studies also use tensegrity to make metamaterials [13,14]. In addition, as for the many demonstrated civil structures, the designs are mainly with regular shapes, where tensegrity units, structure self-similar rules, and linear/circular array patterns can be clearly recognized, such as tensegrity cylindrical triplex [15], V-expander towers [16], tensegrity bridge [17], and torus [18]. Even though the regular shapes can significantly simplify the manufacturing and assembly process, architects may still ask for more, i.e., irregular shapes or shapes based on their innovative ideas, to push their taste of beauty to the finest. Thus, in this paper, to demonstrate the capabilities of

---

[*] Corresponding author

*Email address*: muhaochen@tamu.edu (M. Chen)

tensegrity in achieving the agreement of structural engineering and architectural esthetics at large, we presented a tensegrity-based structure with an irregular shape, a deployable saddle-shaped cable net. The cable net has the advantages of large span, spatial volume, and elegant profile [19-21], which can be used for lightweight covers for roofs, patio shade frames, arenas, stadiums, sports centers, and antennas [22-24], etc. In literature, a few saddle-shaped cable net studies have been investigated. For example, Serdjuks and Rocens studied the displacement reduction of a saddle-shaped cable roof by using a supporting contour in the form of cables, initial flexure, and tie-bars [25]. Su et al. performed the fem static analysis of a saddle-shaped cable net, considering the obvious geometric non-linearity and various loading conditions [26]. Khalkhaliha et al. presented the nonlinear dynamics and vibration control of a saddle-shaped cable net [27]. Li et al. showed a form-finding approach to the equal tension saddle-shaped cable net [28]. Cable net structure has the advantages of lightweight and large span, and there are many successful applications in civil engineering structures [29-31]. However, almost all the existing cable nets are static structures in civil engineering. The lightweight and high-strength properties of cable nets have been exploited thoroughly. The flexibility advantages and efficient control laws of cable nets still need more investigation.

Although the tensegrity paradigm allows one to knit any shape of structures as long as people's imagination and engineering mechanics agree, a significant challenge of the structure is how to deal with the abundant number of strings. In fact, the many strings provide much freedom in control and redundancy to the structure, which might also be a reason that pauses people from moving forward. Because the hardware system to control all the strings is complicated and costly. An intuitive and excellent strategy to deal with this issue is using clustered strings to replace the adjacent strings at the nodes if possible [32]. Therefore, fewer strings, sensors, and actuators are needed [33,34].

Since the clustering strategy may change the behavior of the non-clustered traditional tensegrity, studies are needed on clustering tensegrities. A little research has been conducted on this subject. And as for the control of tensegrity structures, the great challenge has brought much interest from researchers. For example, Veuve demonstrated methods to improve deployment paths and adaptive shape before and after a damage event by a footbridge design [44]. Moored and Bart-Smith studied the cluster strategies and stability analysis of the CTS [35]. Kan et al. demonstrated the rigid body dynamics with sliding cables with an application to the deployment of CTS [36]. Ali et al. presented CTS's static analysis and form-finding problems using a modified dynamic relaxation algorithm [37]. Ma et al. studied the statics and dynamics of a clustered deployable tensegrity cable dome [8]. Kan et al. derived the dynamics of CTS via the positional formulation FEM [38]. Ma et al. proposed a nonlinear shape control law for the nonlinear dynamics of CTS [39]. Moreover, many mechanical cable-pulley systems such as robots [40,41], cranes [42], and parachute systems [43] use flexible and high-strength cables to redirect tensions and balanced forces. But still, the research in deployable cable nets is limited. This paper combines the sliding cables and pulley systems, implements strings as the actuators, and proposes a deployable hyperbolic paraboloid cable net structure. We also present and compare an open-loop and closed-loop control approach to the deployable saddle-shaped cable net with numerical simulations.

The paper is structured as follows. Section 2 gives the statics, nonlinear and linearized clustered tensegrity dynamics equations. Section 3 first shows the topology design of the deployable saddle-shaped cable net. Then, the form-finding, stability, and free vibration analysis are conducted. Section 4 describes the deployment trajectory and performs the open-loop and closed-loop control of the deployable cable net. Section 5 presents an experiment and compares the results with pseudo-static and dynamic simulations. Section 6 summarizes the conclusions.

## 2 The clustered tensegrity dynamics and statics

This section reviews the statics and dynamics of CTS with a nodal coordinate vector as the generalized coordinate. Since the statics is a particular case of dynamics by removing the time derivative terms, we first present the dynamics as follows.

## 2.1 The clustered tensegrity dynamics

The dynamics equation of CTS with constraint is [39]:

$$E_a^T(M\ddot{n} + D\dot{n} + Kn) = E_a^T(f_{ex} - g), \qquad (1)$$

where $E_a \in \mathbb{R}^{3n_n \times n_a}$ is an index matrix to extract the free nodal coordinate $n_a = E_a^T n \in \mathbb{R}^{n_a}$ from the total coordinate vector $n \in \mathbb{R}^{n_n}$, $n_n$ is the number of the total nodes, $n_a$ is the number of free nodes, and $M \in \mathbb{R}^{3n_n \times 3n_n}$ is the mass matrix [39]:

$$M = \frac{1}{6}(|C|^T \widehat{m}|C| + \lfloor|C|^T \widehat{m}|C|\rfloor) \otimes I_3, \qquad (2)$$

where $|V|$ is an operator that gets the absolute value of each element of a given matrix $V$, and the operator $\lfloor V \rfloor$ sets all the off-diagonal to zero. $C \in \mathbb{R}^{n_e \times n_n}$ is the connectivity matrix (consists of a "-1" at the $i$th column, a "+1" at the $j$th column, and zeros elsewhere to define a structure member connecting from $n_i$ to $n_j$). $m = \widehat{\rho}\widehat{A}l_0 \in \mathbb{R}^{n_e}$ is the mass vector of all members, where $\rho \in \mathbb{R}^{n_e}$, $A \in \mathbb{R}^{n_e}$, and $l_0 \in \mathbb{R}^{n_e}$ are the mass density, cross-section area, and rest length vectors of the structure members before clustering, and the $\widehat{v}$ is an operator that converts a vector $v$ into a diagonal matrix.

Note that the mass vector and matrix of the CTS are not constants because the relative motion of clustered strings will change the cable mass distribution in each segment. The damping matrix $D \in \mathbb{R}^{3n_n \times 3n_n}$ is [39]:

$$D = \xi A_{2c} \widehat{d}_c A_{2c}^T, \qquad (3)$$

where $\xi$ is the damping coefficient of the material, $A_{2c} = (C^T \otimes I_3)\text{b.d.}(H)\hat{l}^{-1}S^T \in \mathbb{R}^{3n_n \times n_{ec}}$ is the equilibrium matrix, and $d_c = \frac{2\sqrt{3}}{3}\widehat{\rho}_c^{\frac{1}{2}}\widehat{A}_c E_c^{\frac{1}{2}}$ is the critical damping coefficient of the axial members. $\rho_c \in \mathbb{R}^{n_{ec}}$, $A_c \in \mathbb{R}^{n_{ec}}$, and $E_c \in \mathbb{R}^{n_{ec}}$ are the mass density, cross-section area, and young's modulus vectors of the clustered tensegrity structure. $S \in \mathbb{R}^{n_{ec} \times n_e}$ is the clustering matrix (consists of a "-1" at the $i$th column, a "+1" at the $j$th column, and zeros elsewhere to define a cluster string connecting from $s_i$ to $s_j$) [39]. The tangent stiffness matrix $K \in \mathbb{R}^{3n_n \times 3n_n}$ is:

$$K = \left(C^T \widehat{\hat{l}^{-1}S^T} t_c C\right) \otimes I_3 n, \qquad (4)$$

where $l \in \mathbb{R}^{n_e}$ is the length vector of all members before clustering and $t_c \in \mathbb{R}^{n_e}$ is the force vector of all members after clustering. The dynamics equation Eq.(1) can also be written in terms of free nodal coordinate $n_a$ for the convenience of programming as follows:

$$M_{aa}\ddot{n}_a + D_{aa}\dot{n}_a + K_{aa}n_a = E_a^T f_{ex} - M_{ab}\ddot{n}_b - D_{ab}\dot{n}_b - K_{ab}n_b - E_a^T g, \qquad (5)$$

where $n_b = E_b^T n \in \mathbb{R}^{n_b}$ is the constrained nodal coordinate, $n_b$ is the number of constraints nodes, and $E_b$ is an index matrix of the constraint nodes. $M_{aa}, D_{aa}, K_{aa}$, and $M_{ab}, D_{ab}, K_{ab}$ are respectively:

$$M_{aa} = E_a^T M E_a, M_{ab} = E_a^T M E_b, \qquad (6)$$
$$D_{aa} = E_a^T D E_a, D_{ab} = E_a^T D E_b, \qquad (7)$$
$$D_{aa} = E_a^T D E_a, D_{ab} = E_a^T D E_b. \qquad (8)$$

## 2.2 The clustered tensegrity statics
### 2.2.1 Equilibrium equation

Let the acceleration $\ddot{n}$ and velocity $\dot{n}$ in Eq. (1) be zeros, the dynamics equation will be reduced into the static equilibrium equation in terms of nodal coordinate vector $n$:

$$E_a^T K n = E_a^T (f_{ex} - g), \qquad (9)$$

where $w = f_{ex} - g \in \mathbb{R}^{n_n}$ is the external force, $g = \frac{1}{2}(|C|^T m) \otimes g_0 \in \mathbb{R}^{n_n}$ is the gravity force, and $g_0 = $

$[a_x \ a_y \ a_z]^T \in \mathbb{R}^3$ is the gravity vector. Similarly, we have an equivalent form as Eq. (9) written in terms of the free nodal coordinate $n_a$:

$$K_{aa}n_a = E_a^T f_{ex} + K_{ab}n_b - E_a^T g. \tag{10}$$

The equilibrium equation Eq. (9) can also be written in terms of the force vector $t_c$:

$$E_a^T A_{2c} t_c = E_a^T (f_{ex} - g). \tag{11}$$

#### 2.2.2 Compatibility matrix

For CTS, the compatibility equation denotes the relation between nodal coordinate $n$ and structure member length $l_c$, given as:

$$B_{lc} dn = dl_c, \tag{12}$$

where $B_{lc} \in \mathbb{R}^{n_{ec} \times 3n_n}$ is the compatibility matrix of the CTS:

$$B_{lc} = S\hat{l}^{-1} b.d.(H)^T (C \otimes I_3). \tag{13}$$

where $b.d.(H)$ is the block diagonal matrix of $H = NC^T$. Noted that the compatibility and equilibrium matrix of the CTS has the following relationship: $B_{lc}{}^T = A_{2c}$.

### 2.3 Linearized dynamics

Take the total derivative of Eq. (1) and keep the linear terms, one can have the linearized dynamics:

$$E_a^T (M dn + D dn + K_T dn + K_{l_{0c}} dl_{0c}) = E_a^T df_{ex}, \tag{14}$$

where $K_T \in \mathbb{R}^{3n_n \times 3n_n}$ is the tangent stiffness matrix given in Eq. (17). And $K_{l_{0c}} \in \mathbb{R}^{3n_n \times n_e}$ is the sensitivity matrix, which denotes the rest length with respect to the nodal force, given in Eq.(18). And Eq. (14) has the following equivalent form where the free and constraint nodes are separated on the two sides:

$$M_{aa} d\ddot{n}_a + D_{aa} d\dot{n}_a + K_{T_{aa}} dn_a \\ = E_a^T df_{ex} - E_a^T K_{l_{0c}} dl_{0c} - M_{ab} d\ddot{n}_b - D_{ab} d\dot{n}_b - K_{T_{ab}} dn_b, \tag{15}$$

where:

$$K_{T_{aa}} = E_a^T K_T E_a, K_{T_{ab}} = E_a^T K_T E_b. \tag{16}$$

#### 2.3.1 Tangent stiffness matrix

The tangent stiffness matrix is [39]:

$$K_T = \frac{\partial (Kn)}{\partial n^T} = \left(C^T \hat{l}^{-1} \widehat{S^T t_c} C\right) \otimes I_3 + A_{2c} \widehat{E}_{tc} \widehat{A}_c \hat{l}_{0c}^{-1} A_{2c}^T - A_2 \widehat{S^T t_c} \hat{l}^{-1} A_2^T. \tag{17}$$

The first and the third part of Eq. (17) is usually called the geometry stiffness matrix $K_G = \left(C^T \widehat{l^{-1} S^T} t_c C\right) \otimes I_3 - A_2 \widehat{S^T t_c} \hat{l}^{-1} A_2^T$ which is determined by structure topology and member forces. The second part is called the material stiffness $K_E = A_{2c} \widehat{E}_{tc} \widehat{A}_c \hat{l}_c^{-1} A_{2c}^T$ which is governed by the structure configuration and structure elements' axial stiffness.

#### 2.3.2 Sensitivity matrix

The sensitivity matrix of rest length with respect to the nodal force can be written as:

$$K_{l_{0c}} = \frac{\partial (Kn)}{\partial l_{0c}^T} = -A_{2c} \widehat{E}_{tc} \widehat{A}_c \hat{l}_c \hat{l}_{0c}^{-2}. \tag{18}$$

We also define two useful sensitivity matrices: $K_{t_c, l_{0c}}$ and $K_{t_c, w}$ to denote the member length to member force and external force to member force, respectively. The derivation can be achieved based on the linearized statics equations as follows. Let the cross-sectional area $A_c$ of members be a constant. Since the equilibrium is a function of free nodal coordinate $n_a$ and rest length $l_{0c}$, the total derivate of member force is:

$$dt_c = \frac{\partial t_c}{\partial n} E_a dn_a + \frac{\partial t_c}{\partial l_{0c}} dl_{0c}. \tag{19}$$

From the above equation, we can observe that the difference in member force is influenced by the free nodal coordinate $n_a$ and rest length $l_{0c}$. But $n_a$ and $l_{0c}$ are not independent, and they are related by the equilibrium equation Eq. (9). The total derivate of Eq. (9) is:

$$K_{Taa}dn_a + E_a^T K_{l_{oc}} dl_{0c} = E_a^T dw. \tag{20}$$

The two terms $\frac{\partial t_c}{\partial n}$ and $\frac{\partial t_c}{\partial l_{0c}}$ in Eq.(19) are respectively:

$$\frac{\partial t_c}{\partial n} = \frac{\partial t_c}{\partial l_c}\frac{\partial l_c}{\partial n} = \widehat{E}_{tc}\widehat{A}_c \hat{l}_{0c}^{-1} A_{2c}^T, \tag{21}$$

$$\frac{\partial t_c}{\partial l_{0c}} = -\widehat{E}_{tc}\widehat{A}_c \hat{l}_c \hat{l}_{0c}^{-2}. \tag{22}$$

From Eq.(20), we have:

$$dn_a = K_{Taa}^{-1} E_a^T (dw - K_{l_{oc}} dl_{0c}). \tag{23}$$

The sensitivity of member length to free nodal coordinate $K_{n_a,l_{oc}}$ can be obtained as:

$$K_{n_a,l_{oc}} = -K_{Taa}^{-1} E_a^T K_{l_{oc}}. \tag{24}$$

The sensitivity of external force to free nodal coordinate $K_{n_a,w}$ is:

$$K_{n_a,w} = K_{Taa}^{-1} E_a^T. \tag{25}$$

Substitute Eqs.(21), (22), and (23) into Eq.(19), we have:

$$dt_c = \widehat{E}_{tc}\widehat{A}_c \hat{l}_{0c}^{-1} A_{2c}^T E_a K_{n_a,w} dw + \widehat{E}_{tc}\widehat{A}_c \hat{l}_{0c}^{-1}(A_{2c}^T E_a K_{n_a,l_{oc}} - \hat{l}_c \hat{l}_{0c}^{-1})dl_{0c}. \tag{26}$$

From the above equation, the sensitivity matrix of member length to member force $K_{t_c,l_{oc}}$ is:

$$K_{t_c,l_{oc}} = \widehat{E}_{tc}\widehat{A}_c \hat{l}_{0c}^{-1}(A_{2c}^T E_a K_{n_a,l_{oc}} - \hat{l}_c \hat{l}_{0c}^{-1}). \tag{27}$$

The sensitivity matrix of external force to member force $K_{t_c,w}$ is:

$$K_{t_c,w} = \widehat{E}_{tc}\widehat{A}_c \hat{l}_{0c}^{-1} A_{2c}^T E_a K_{n_a,w}. \tag{28}$$

### 2.3.3 Modal analysis

For tensegrity dynamics with constraints, the free vibration response can be obtained from Eq. (15) by neglecting damping force, external force, change of rest length, and motion of boundary nodes:

$$M_{aa} d\ddot{n}_a + K_{Taa} dn_a = 0. \tag{29}$$

Substituting $dn_a = \varphi sin(\omega t - \theta)$ into the above equation, the free vibration mode can be solved by a generalized eigenvalue problem:

$$K_{Taa}\varphi = \omega^2 M_{aa}\varphi, \tag{30}$$

where $\omega$ is the natural frequency of the system and $\varphi$ is the corresponding eigenvector representing the mode shapes.

## 3 Design of a deployable hyperbolic paraboloid cable net structure

### 3.1 Topology and arrangement of the clustered cables

As shown in Figure 1, the movable pulleys, fixed pulleys, and actuators from the mechanical cable-pulley system are combined into the hyperbolic paraboloid cable net. The actuators are used to change the rest length and prestress of the clustered cable. The fixed pulleys connect the clustered cables to the structure boundary. The movable pulleys connect the clustered cables. The bottom part is the structural boundary of the cable net, in which the red and black lines are clustered cables and structural boundaries. The cables in the same color are clustered cables whose rest length can be changed by the actuators.

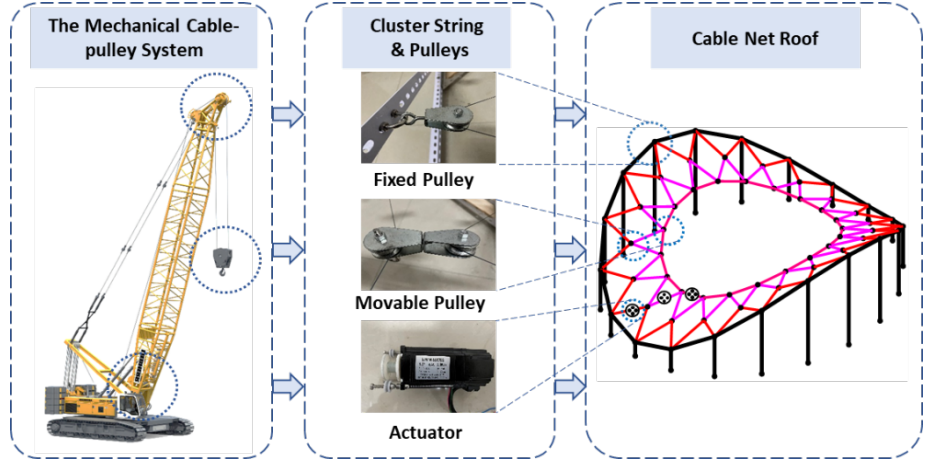

Figure 1 Configuration of the clustered cable-actuated deployable hyperbolic paraboloid cable net.

The geometry and topology of the cable net are shown in Figure 2. The boundary nodes of the cable net structure are located on an ellipse from the top view in Figure 2(a). The boundary nodes are evenly distributed on the ellipse with a total number of $p$, also called as the complexity of the structure. The equation of the ellipse is:

$$\frac{x^2}{R_x^2} + \frac{y^2}{R_y^2} = 1, \tag{31}$$

where $R_x$ and $R_y$ are respectively the major radius and minor radius of the ellipse.

The cable net is composed of $q$ diagonal cables and one hoop cable at the center, where we call $q$ the complexity of the cable net in the radial direction. The cable net has two types of nodes, which are $p$ boundary pinned nodes and $p(q-1)$ free nodes. The pulleys at the nodes connect the individual cables into clustered cables. Each free node is composed of two pulleys connecting diagonal and hoop cables, respectively. In Figure 2, we set $p = 20$ and $q = 2$ as an example for illustration.

The X and Y coordinates of the inner free nodes on the inner hoop cable is scaled by $c$ times of the boundary pinned nodes, as shown in Figure 2(a), where $c$ is a coefficient of the deployment ratio. The node of the cable net is at the intersection of the hyperbolic paraboloid and the elliptical cylinder, as shown in Figure 2(b). The equation of the hyperbolic paraboloid is:

$$\frac{y^2}{b^2} - \frac{x^2}{a^2} = z, \tag{32}$$

where a and b are constants that dictate the level of curvature in the xz and yz planes, respectively. The vertical distance between the top node and bottom node on the boundary is $R_y^2/b^2 + R_x^2/a^2$, as shown in Figure 2(c,d). The topology and geometry of the deployable hyperbolic paraboloid cable net can be simply parameterized by six variables $p$, $q$, $R_x$, $R_y$, $a$, $b$ and $c$.

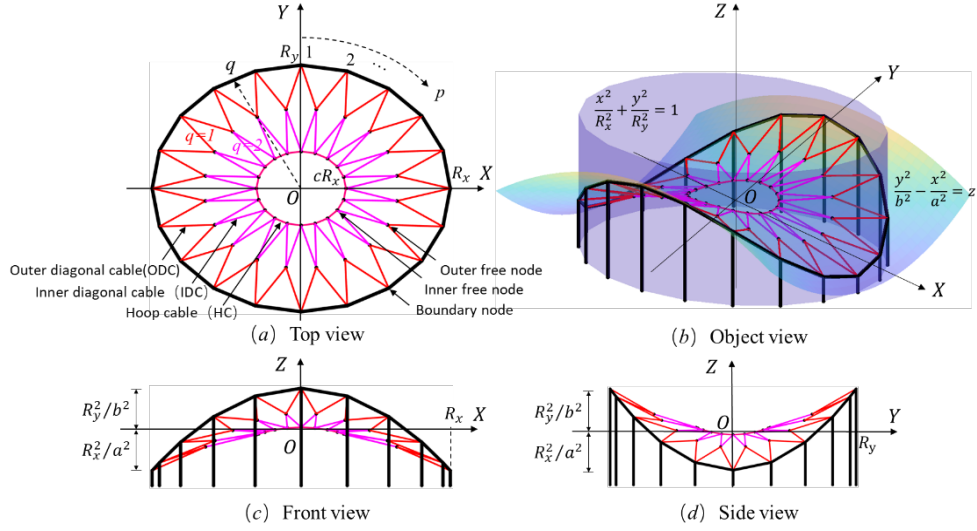

Figure 2 Topology and initial configuration of the deployable hyperbolic paraboloid cable net.

We can adjust the deployment ratio of the structure by changing the rest length of the hoop and diagonal cables simultaneously. By increasing the rest length of the hoop cable and decreasing the rest length of diagonal cables, the cable net can be deployed. And by decreasing the rest length of the hoop cable and increasing the rest length of the diagonal cables, the cable net can be folded. The deployment trajectory of the hyperbolic paraboloid cable net is shown in Figure 3.

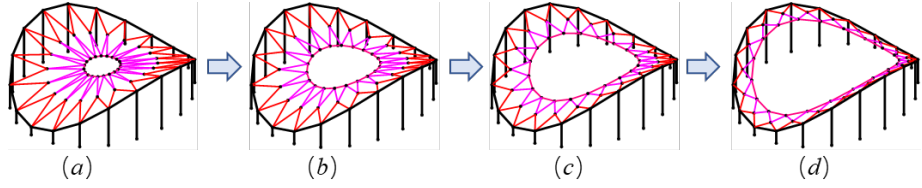

Figure 3 Deployment trajectory of the hyperbolic paraboloid cable net.

### 3.2 Form-finding of the cable net

The configuration of the deployable hyperbolic paraboloid cable net structure is complex and coupled with the prestress design, so the equilibrium configuration of the cable net structure cannot be arbitrarily given. The configuration given in the previous section is the initial configuration. The form-finding method needs to recalculate the configuration with an integral feasible prestress mode [45] around the initial configuration. In this section, the form-finding of the deployable cable net is obtained by solving nonlinear equilibrium based on the Newton Raphson method. The flowchart of the form-finding algorithm is shown in Figure 4. The form-finding process of the deployable cable net is as follows:

1) Input: initial configuration $\boldsymbol{n}^0$, connectivity matrix $\boldsymbol{C}$, clustering matrix $\boldsymbol{S}$, rest length $\boldsymbol{l}_{0c}$, axial stiffness $\boldsymbol{E}_c \boldsymbol{A}_c$ $i = 0$.
2) Calculate the unbalanced force $\boldsymbol{F}_p^i$.
3) Calculate the tangent stiffness $\boldsymbol{K}_{Taa}|_{\boldsymbol{n}^i}$.
4) Calculate the difference between free nodes $\mathrm{d}\boldsymbol{n}_a^i$.
5) Renew the nodal coordinate $\boldsymbol{n}_a^{i+1}$.
6) Judge the equilibrium criteria. If satisfied, output, if not, set $i = i + 1$, go to step 2).

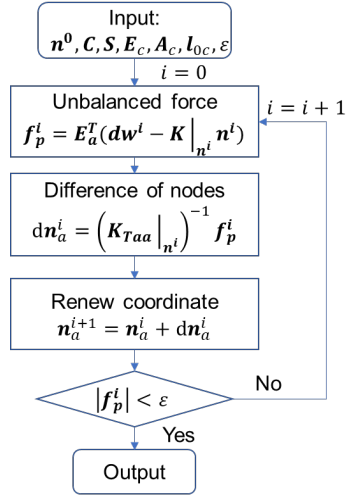

Figure 4 Form-finding procedure of the hyperbolic paraboloid cable net.

The cable net structure, parameterized by the seven variables in Figure 2, is given as the initial configuration. Before the loading condition design, we first assign one tension to all the cables, i.e., $10^4\text{N}$, to initialize the prestress design process. We should point out that the prestress would be redesigned based on the feasible loading conditions in the next section. The cross-sectional of the strings and bars are designed by the minimal mass design under yielding and buckling constraints of strings and bars [18,46,47], and the safety coefficient is set as 0.1. The equilibrium configuration of the hyperbolic paraboloid cable net is obtained by solving the nonlinear equilibrium equation, as shown in Eq.(9). The equilibrium configuration is plotted in the solid line, and the initial configuration is plotted in a dotted line, as shown in Figure 5. We can observe that the equilibrium configuration is slightly different from the initial configuration.

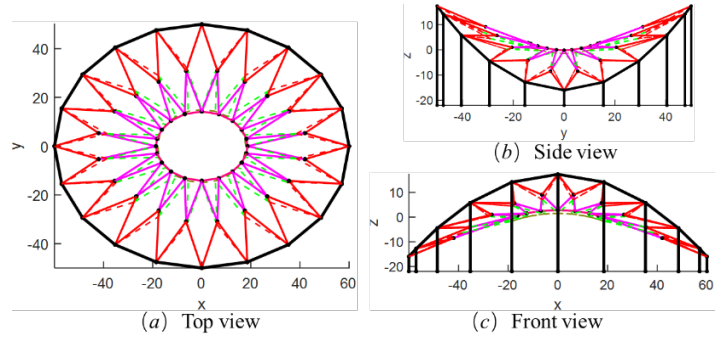

Figure 5 Comparison of initial and equilibrium configuration of the hyperbolic paraboloid cable net

### 3.3 Prestress design of the cable net

The equilibrium result obtained by the form-finding method in the previous section does not guarantee the prestress of the cable net is applicable, i.e., the force in the structure members may exceed their yield strength, the limitations of the hardware, or one wants to simplify the deployment process. In such cases, we must redesign the prestress. Note that the configuration of the form-finding results in Section 3.2 is in equilibrium, which means the configuration has at least one integral feasible prestress mode. According to Eq.(11), the equilibrium equation can be simplified to:

$$\overline{A}_{2c} t_c = w_a, \qquad (33)$$

where $w_a$ is the external force exerted on the free nodes. $\overline{A}_{2c} \in R^{n_a \times n_{ec}}$ is the equilibrium matrix with boundary constraint:

$$\overline{A}_{2c} = E_a^T A_{2c}. \qquad (34)$$

The singular value decomposition[48] reveals the four subspaces of $\bar{A}_{2c}$:

$$\bar{A}_{2c} = U\Sigma V^{\mathrm{T}} = [U_1 \quad U_2]\begin{bmatrix} \Sigma_0 & 0 \\ 0 & 0 \end{bmatrix}\begin{bmatrix} V_1^{\mathrm{T}} \\ V_2^{\mathrm{T}} \end{bmatrix}, \tag{35}$$

where $U \in R^{n_a \times n_a}$, and $V \in R^{n_{ec} \times n_{ec}}$ are orthogonal matrices. Let $r = \text{rank}(\bar{A}_{2c})$ be the rank of $\bar{A}_{2c}$. $V_1 \in R^{n_{ec} \times r}$, $V_2 \in R^{n_{ec} \times (n_{ec}-r)}$ is respectively the row space and null space of $A_r$, and $U_1 \in R^{n_a \times r}$, $U_2 \in R^{n_a \times (n_a-r)}$ is respectively the column space and left null space of $\bar{A}_{2c}$. $\bar{A}_{2c}V_2 = 0$ and $\bar{A}_{2c}^T U_2 = 0$, $V_2$ and $U_2$ are the self-stress mode and mechanism mode of the tensegrity structure, respectively. The solution of Eq.(33) is:

$$t_c = (\bar{A}_{2c})^+ w_a + V_2 z, \tag{36}$$

where the first part is the particular solution corresponding to external force and the second part is the general solution corresponding to the prestress. The arbitrary vector $z$ is the prestress coefficient vector that combines the columns in prestress mode $V_2$. The matrix $e_d$ is used to extract the force of certain designed members $t_d$:

$$t_d = e_d^{\mathrm{T}} t_c. \tag{37}$$

To simplify the deployment process with feasible tensions, we can assign the force to the specific structure members $t_d$ to be a designed force $\bar{t}_d$. For example, we assign the force of ODC to be $\bar{t}_d = 10^4 \text{N}$. Noted that the tension of cable is about $10^4 \text{N}$ to $10^5 \text{N}$ in civil engineering practice of such a scale model, but for real cases, the accurate value of cable force needs to satisfy yielding, buckling constraints, natural frequency, stiffness, and many other design constraints. Usually, the number of members that can be designed to a specific force is identical to the number of integral feasible prestress mode $n_{ec} - \text{r}$, we have:

$$\bar{t}_d = e_d^T [(\bar{A}_{2c})^+ w_a + V_2 z]. \tag{38}$$

The solution to the above equation is:

$$z = (e_d^T V_2)^{-1} [\bar{t}_d - e_d^T (\bar{A}_{2c})^+ w_a]. \tag{39}$$

Substituting the solution of $z$ from Eq. (39) into Eq.(36), we can compute the forces in the cable net. It should be noted that for the proposed deployable cable net structure, which has two clustered cables connected to the free nodes, $w_a = 0$ is necessary for the feasibility of the structure. If $w_a \neq 0$, the two clustered cables connected to the node with external force will deform to a configuration with no prestress modes, which means the prestress of the structure cannot be redesigned. In this study, we find that the deployable cable net structure has only one prestress mode, so $z$ is an arbitrary scalar instead of a vector.

### 3.4 Stability and free vibration analysis

Using clustered cables and pulleys will generally decrease the stiffness of the cable net structure [48]. We must study the stability and natural frequency of the deployable hyperbolic paraboloid cable net structure to ensure the structure configuration is feasible for engineering applications.

The essence of the stability and natural frequency analysis is to perform the eigenvalue analysis and generalized eigenvalue analysis of the tangent stiffness matrix of Eq.(17). The free vibration mode shapes with the first four frequencies are shown in Figure 6, where the solid line is the free vibration mode, and the dotted line is the equilibrium configuration. The first mode shape corresponds to the displacement along the X-axis, and the frequency is 0.4034Hz. The second mode shape is the displacement along the Y-axis, and the frequency is 0.4045Hz. The third mode shape denotes the vertical displacement, and the frequency is 0.4720Hz. The fourth mode shape reveals the local relative motion of the pulley and sliding cable, and the frequency is 0.4825Hz.

The eigenvalue analysis of the tangent stiffness reveals the stiffness distribution and stability of the structure. The first four eigenvalues of the tangent stiffness matrix are 158.2586, 158.2700, 158.2727, 158.3129N/m, and the mode shapes of the first four modes are all local relative motion of pulley and sliding cables.

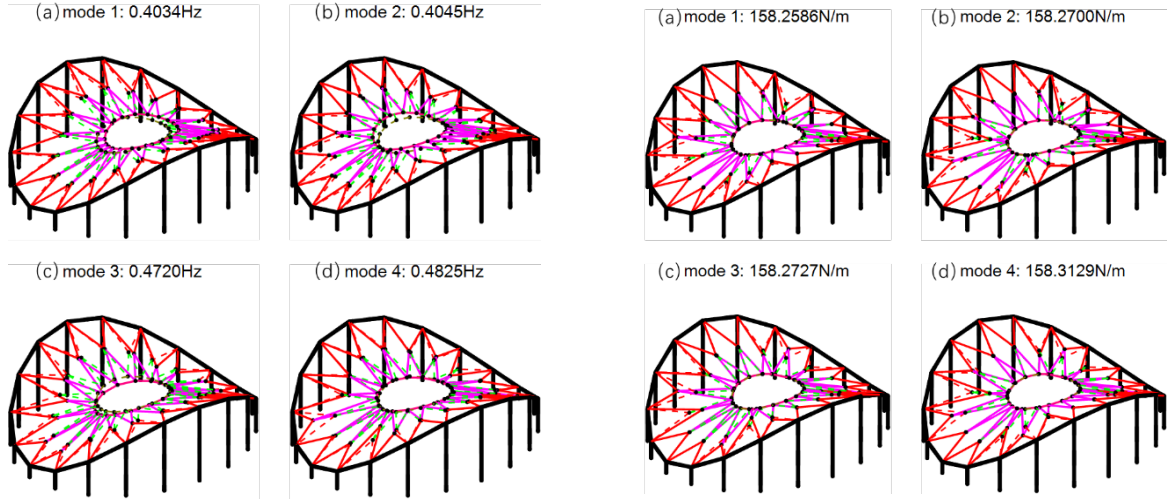

Figure 6 The first four free vibration frequencies and mode shapes.

Figure 7 The first four minimal stiffness eigenvalues and mode shapes.

## 4  Shape control of the deployable cable net

### 4.1  Deployment trajectory design

Unlike the traditional retractable roof structures made of rigid elements, the moving trajectories of the rigid parts are often easy to set up. But it has always been a complex problem to control the flexible structures with prestress requirements, i.e., the cable net structure. To make a precise control, it is better to assign a trajectory as a tracking reference for the deployment process. Since we would like to use less control energy to achieve a smooth control while keeping the structure stable during the transition, it is better to switch the equilibrium states between the initial and final configurations. Thus, we perform the statics form-finding simulation and store the equilibrium configurations as the objective trajectory. To demonstrate this approach, we keep the cross-sectional area and rest length of hoop cables constant and gradually change the rest length of diagonal cables. The deployment trajectory design is shown in Figure 8.

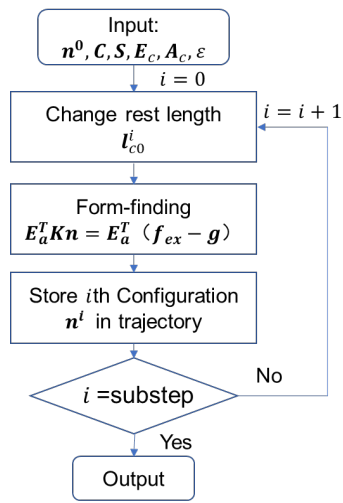

Figure 8 Flow chart of the deployment trajectory design.

In this study, we reduce the rest length of ODC and IDC by 700m in 10 substeps and note that the rest length of ODC,

IDC, and HC are respectively 866.27, 799.30, 103.60m. The equilibrium configuration in the 10 substeps is solved by the form-finding method. The configuration of the deployment trajectory is shown in Figure 9, where node 3 is the inner free node on the positive X-axis, and the tension of the cables in the deployment process is shown in Figure 10. We can observe that the cable net deploys smoothly on the hyperbolic paraboloid, so the deployment trajectory is reasonable. But the tension in cables increases more than 1,000 times in the deployment process, which is difficult to accept in engineering practice. So, apart from designing a smooth trajectory, we also need to consider the feasibility of the deployment process. That is, design prestress of the structure to keep the tensions within a small variation range during the whole deployment process.

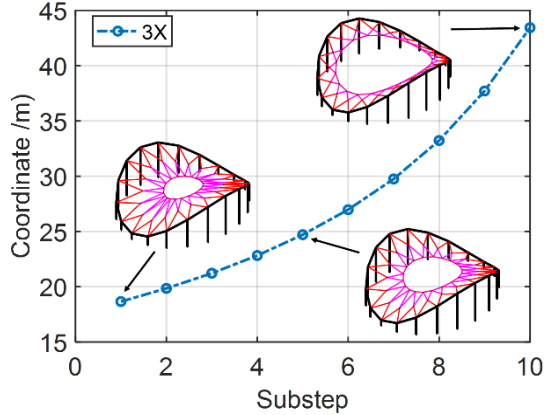

Figure 9 Deployment trajectory and configurations of the cable net.

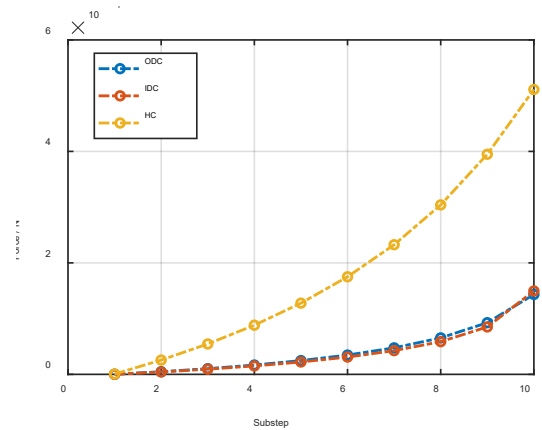

Figure 10 The tensions among the members during the deployment process.

## 4.2 Redesign prestress in the deployment process

Since the prestress in Figure 10 is too large to implement in engineering practice, in this section, we present the redesign of prestress for the deployable cable net by the method given in Section 3.3.

We assign the force of ODC to be $\bar{t}_d = 10^4 \text{N}$ in the deployment process and recalculate the prestress of the structure by Eq. (39) and Eq.(36). The results of the forces in the strings during the deployment trajectory are shown in Figure 11. We can observe that the force of ODC is identical to the designed force, and the prestress design result is more feasible than the force in Figure 10 after the prestress redesign.

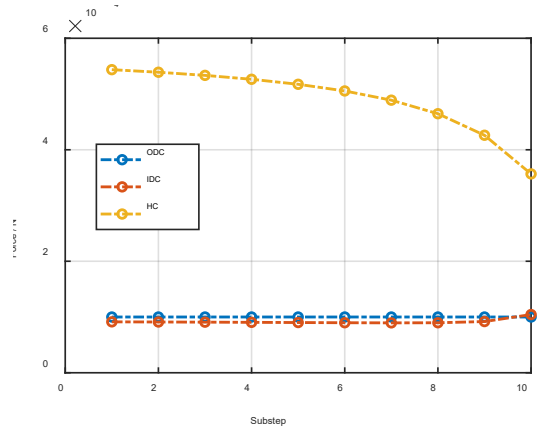

Figure 11 The forces in the strings after the prestress design during the deployment process.

## 4.3 Open-loop control strategy

From the previous section of the trajectory design, we can compute the rest length of the cables, as the control variable, at each substep of the deployment trajectory and directly use it for the open-loop control. The block diagram of open-loop control is shown in Figure 12.

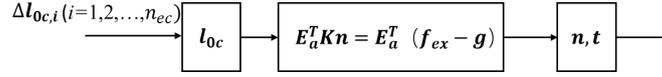

Figure 12 Block diagram of open-loop control.

The rest length of members in each substep of the deployment trajectory can be calculated by the following equation:

$$l_{0c} = (\hat{t}_c + \hat{E}_c \hat{A}_c)^{-1} \hat{E}_c \hat{A}_c l_c. \tag{40}$$

Substitute the prestress design result $t_c$ from Figure 11 and the deployment trajectory result $l_c$ from Figure 9 into the above equation, we can get the results of the rest length in the deployment trajectory, as shown in Figure 13. Substitute the $l_{0c}$ into Eq. (9) and perform the form-finding analysis, we can obtain the pseudo-static deployment process.

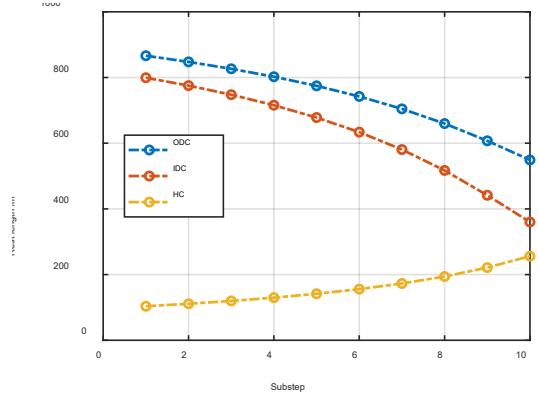

Figure 13 The rest length of the cables during the deployment process.

### 4.4 Closed-loop control strategy

The open-loop control strategy assumes the deployment is slow-motion so that each substep can be viewed as a static process, and the actuation and structure model is accurate. But for an actual application, one cannot guarantee the deployment model is as precise as the mathematical model. For example, the rest length of the strings may be influenced by the pulley friction, pulley radius variation by the number of cable rolls, material creep, etc. Often, an open-loop control may not be enough, and closed-loop control law is needed.

Note that on the one hand, due to the high axial stiffness of the strings, the sensitivity of member rest length to member force $K_{t_c,l_{oc}}$ is relatively high. That is, any small error in the rest length will result in a significant member force. On the other hand, from Eq.(27), we can see that the sensitivity of member rest length to free nodal coordinate $K_{n_a,l_{oc}}$ is much lower than the sensitivity of member rest length to member force $K_{t_c,l_{oc}}$. That means the influence of the free nodal coordinate is much smaller than the member force caused by the error of rest length. So, during the deployment process, it is vital to constrain the error of members' force. If the error of member force is small, the nodal coordinate can be controlled automatically.

Based on the above analysis, we propose a closed-loop control strategy, as shown in Figure 14. We measure the force of the hoop cable $t_1$ and compared with the target value $\bar{t}_1$, the error $\Delta t_1 = \bar{t}_1 - t_1$ is used to calculate the control value of the rest length:

$$\Delta l_{0c,1} = [e_1^T K_{t,l_{oc}} e_1]^{-1} \Delta t_1, \tag{41}$$

where $e_1$ is the vector to extract the value of the hoop cable $t_1 = e_1^T t_c$. $e_1^T K_{t,l_{oc}} e_1$ denotes the sensitivity of the hoop cable's rest length to its member force. The change of rest length of the diagonal cable $\Delta l_{0c,i} (i = 2,3,\cdots n_{ec})$, given in Eq. (40), is applied to the system to control the deployment ratio of the hyperbolic paraboloid cable net. So, in the closed-loop control, the rest length of the diagonal cable is directly given to control the structural configuration, while the hoop cable is used to control the prestress of the structure.

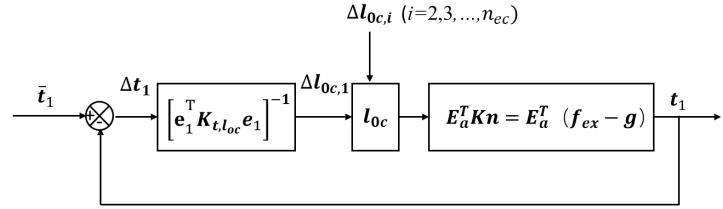

Figure 14 Block diagram of the closed-loop control strategy.

## 5 Simulation and experimental results

### 5.1 The experiment model

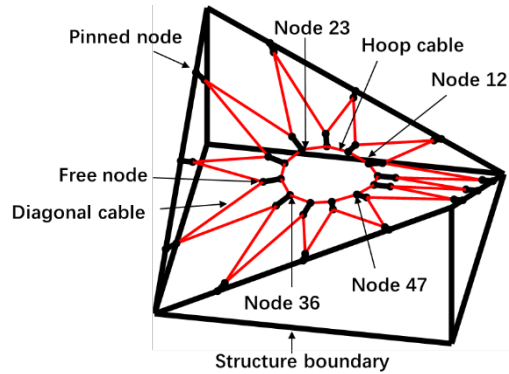

Figure 15 Design of the deployable cable net experimental model.

The experimental model of the deployable cable net is shown in Figure 15. For the simplicity of the experiment, the structure boundary nodes lie on the edges of the straight bar. The experiment structure is 1.05m in length, 1.05m in width, and 0.5m in height. The radius of the inner hoop in the initial configuration is 150mm. Structure complexity is $p = 12$ and $q = 1$ (there is one diagonal cable). Most existing research neglect the pulley size in clustered tensegrity design [35,37,38]. However, the pulley size in 13mm radius and the distance between the two pulleys in the free nodes cannot be ignored for a practical engineering problem. In this simulation, the free and fixed pulleys are respectively modeled by 60mm and 40mm steel bars with $1mm^2$ cross-sectional area. The weight of the free and fixed pulleys is 1.6N and 0.5N.

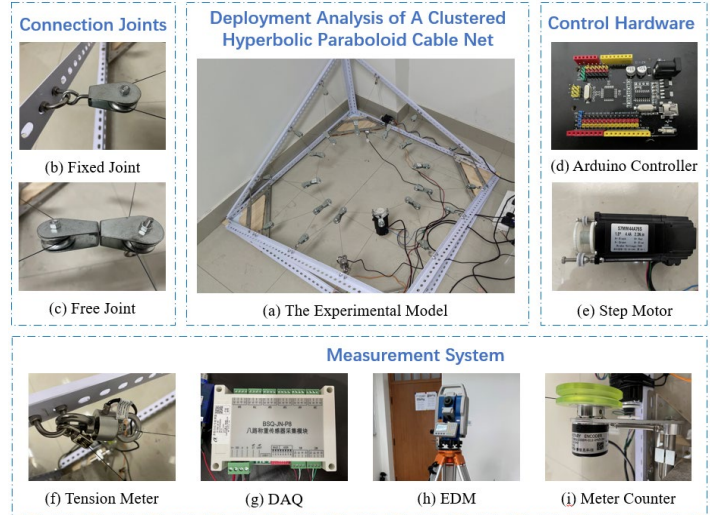

Figure 16 The experimental model of the deployable cable net.

As shown in Figure 16(a), the structure boundary of the experimental model is made of 40*40*2.5mm angle iron. The cable is braided nylon rope with a diameter of 0.6mm, and the product of Young's modulus and cross-sectional area of the nylon rope is $EA = 3,000$N, the material density is $1,150$kg/m$^3$. In the joint design, two pulleys are arranged in one free joint to connect the hoop and diagonal cables. As shown in Figure 16(c), the two pulleys are connected in the end part, and relative rotation motion is allowed along the line connecting the two pulley centers. This is an essential device for the compatibility of the clustered hoop and diagonal cables in a different plane and ensures the equilibrium condition of each pulley. As shown in Figure 16(b), the pinned joint is composed of a semi-circle rigidly connected to the structural base, and the pulley allows rotation along the line connecting the end node and center of the pulley. The pinned joint has three degrees of rotational freedom.

As shown in Figure 16(e), the 42BYGH47 step motor controls the rest length of clustered cables. There is one step motor to control the rest length of hoop cables and two-step motors arranged in the opposite position to control the diagonal cables. As shown in Figure 16(i), the SCN-JK76 meter counter is placed near the step motor for diagonal cables and is used to measure the change of the rest length of diagonal cables actuated by the step motor. Noted that the radius of the step motor changes with the number of cables twisted, so a meter counter is necessary to measure the distance independently. The step motor is controlled by a DM542 micro-step driver and an Arduino controller in Figure 16(d) to change the rotation angle and direction. As shown in Figure 16(f), two tension sensors are placed in the two clustered cables. The digital analogy converter sends the tension data into the Arduino controller for feedback control. The STONEX R1 total station, as shown in Figure 16(h), is used to measure the nodal coordinates of the structure.

### 5.2 Pseudo static and dynamic analysis of the structural deployment

In Section 4, the shape control of the deployable cable net structure is based on the pseudo-static assumption, in which the inertial force and the damping force are ignored. This is accurate when the actuation is at a relatively slow speed so that each step can be viewed as a static process. In this section, we analyze the dynamic response of the deployment process at different speeds and compare it with the pseudo-static deployment results.

In the dynamics analysis of the deployment process, the Runge-Kutta method is used to solve the dynamics equations. The shape control strategy is kept the same for the pseudo-static and dynamic deployment analysis. In dynamics analysis, the change of rest length is exerted on the structure evenly in a certain period of time, as shown in Figure 17. It can be concluded that the longer the control process takes, the closer the dynamics response is to the pseudo-static analysis. Generally, the dynamics time history of member force is close to statics solution if the deployment process is longer than 5 seconds.

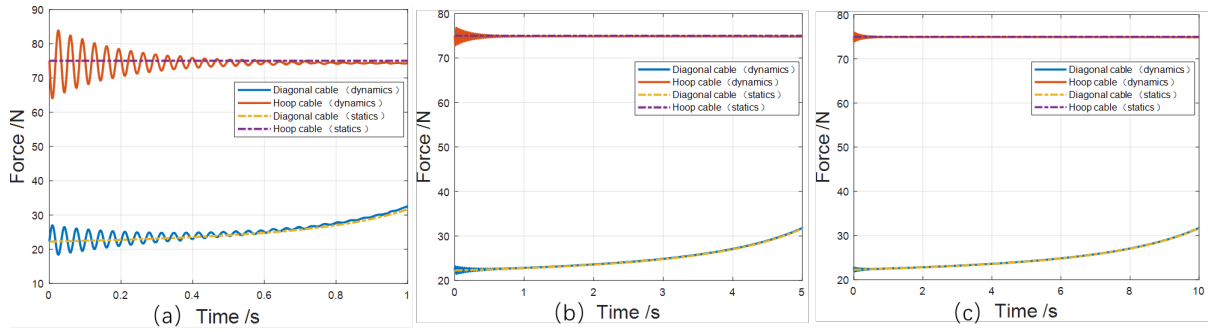

Figure 17 Comparison of the statics and dynamics response of the deployment process.

### 5.3 Experimental study of the deployment

This section compares the experimental study of deploying the clustered cable-actuated cable net structure controlled by open-loop and closed-loop methods. Both the open-loop and closed-loop control are conducted at a very low speed to satisfy the pseudo-static assumption. In the structural deployment, nodal coordinate and member forces are measured to evaluate the accuracy and efficiency of the control methods. Since the structure is symmetric, the X and Y coordinate of nodes 12, 24, 36, and 48 are measured by the STONEX R1 total station, as shown in Figure 16.

#### 5.3.1  Open-loop control

The basic idea of the open-loop control is directly changing the rest length of the clustered cable to control the configuration of the cable net structure without feedback. To achieve the pseudo-static deployment, the control strategy in Section 4.3 is divided into 10 substeps so that the configuration and prestress can be controlled more precisely. The cable net will gradually deploy as planned by changing the rest length of clustered cables according to the control strategy at a very low speed. Figure 18 and Figure 19 are, respectively, the X and Y nodal coordinates of four nodes in the deployment process. Figure 20 shows the cable tensions in the deployment process. Figure 21(a)-(d) shows the configuration of the cable net during the deployment process. The red number in the upper right of the pictures is the reading of the meter counter, which measures the change of the rest length of the diagonal cable.

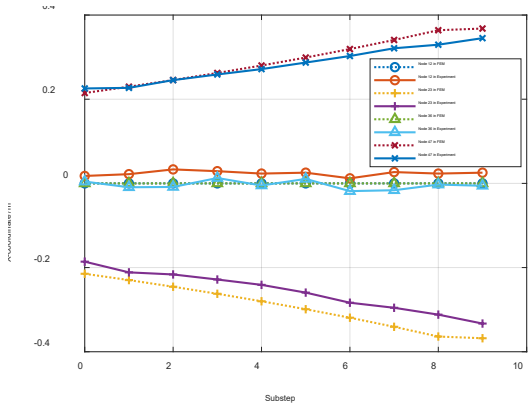 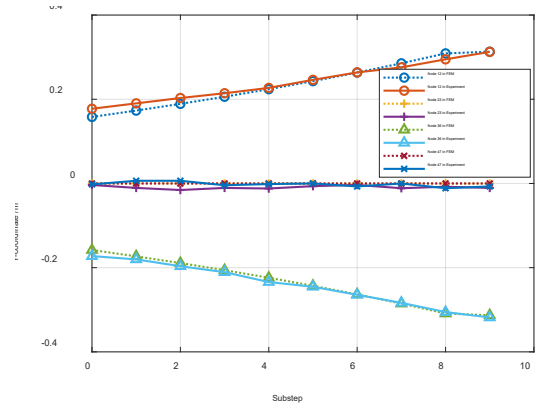

Figure 18 Numerical and experimental results of the X nodal coordinates in the deployment process.

Figure 19 Numerical and experimental results of the Y nodal coordinates in the deployment process.

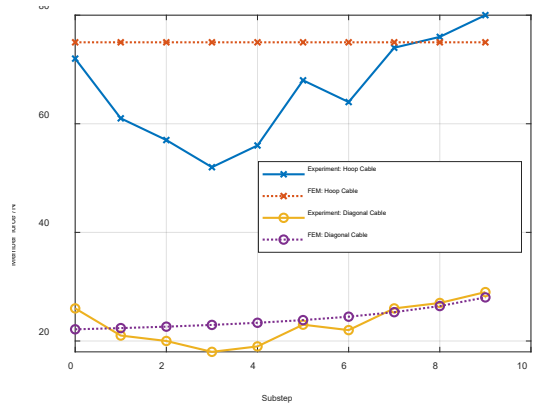
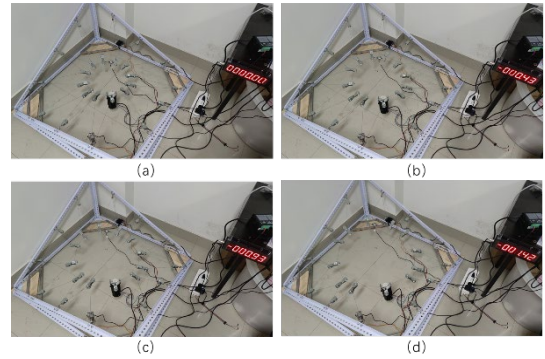

| Figure 20 Numerical and experimental results of the tensions in the cables during the deployment process. | Figure 21 Configuration of the cable net during the deployment process. |

We can observe that both the X and Y nodal coordinate of the experimental result is very close to the simulation result. But the cables' tension has a big difference between the experiment and simulation result. The tension in the hoop and diagonal cables decreases quickly in the first three substeps of the deployment process. However, the tension increases in the following substeps and exceeds the simulation result in the deployed configuration. The small error in result length may not lead to a significant error in configuration but may result in a big error in strain and stress of cables. However, it is reasonable to have this experimental result since the nodal configuration result is more precise than the cables' tension result.

Many factors may influence the accuracy of the controlled value of rest length. Firstly, in the controlling code, the radius of the rod in the step motor is set to be constant. However, in the experiment model, the radius of the rod changes if the number of cables twisted on the rod changes, which can influence the accuracy of the rest length. Secondly, the initial configuration of the open-loop control is not precisely consistent with the simulation result, leading to errors in the deployment process. Thirdly, the friction in the pulley can also influence the cables' tension. A cable net structure is a flexible structure whose prestress plays an essential role in maintaining the stiffness and stability of the structure. Typically, a 40% error in cable tensions is not acceptable in engineering practice. Since cable-net structures are very sensitive to rest length to cable forces, the open-loop control may not satisfy the tolerance of tension errors. A closed-loop control method can reduce the error of cable tensions.

5.3.2    Closed-loop control

In the closed-loop control, the hoop cable force is used as the feedback to control the forces in the hoop cable to be the designed value. In the closed-loop shape control process, the rest length of the diagonal cable is directly used to control the structure configuration, and the rest length of the hoop cable is implemented to control the prestress level of the structure by the closed-loop controller.

The X and Y coordinates and the cable tensions in the deployment process by the closed-loop control are shown in Figure 22Figure 24Figure . We can observe that the nodal coordinate of the experimental result is close to the simulation result. Compare Figure 20 with Figure 24, the cable tensions in the closed-loop control are much closer to the simulation result than in the open-loop one.

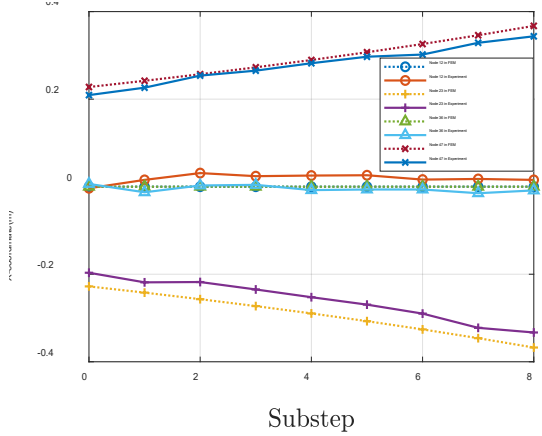

Figure 22 X nodal coordinates in the deployment process.

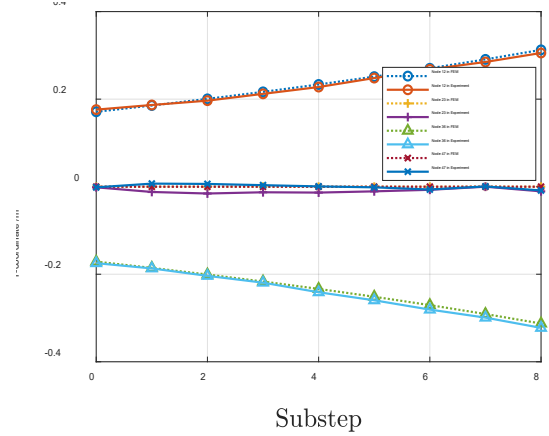

Figure 23 Y nodal coordinates in the deployment process.

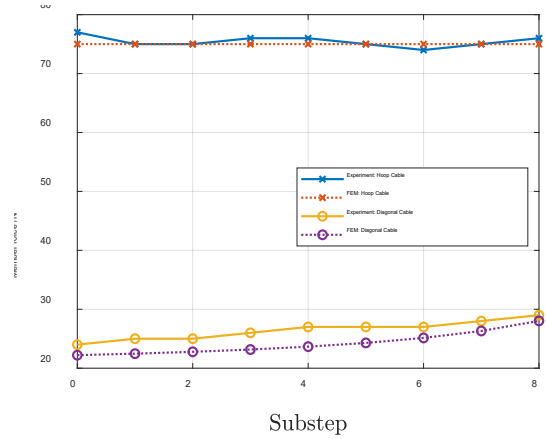

Figure 24 Tension in the cables in the deployment process.

We can conclude that both open-loop and closed-loop control methods can achieve the morphing objectives with satisfactory precision. In the closed-loop control, the cable tensions are around the desired level, while the cable tensions in open-loop control experience a more significant error. So, the closed-loop control performance is much better in controlling the prestress of the deployable cable net than the open-loop control. Since the prestress is vital to the structural stiffness and stability of cable net structure, it is necessary to use closed-loop control in real applications of the cable nets. Besides the deployment control of the cable net, the performance of the cable net in service also deserves concern. For example, the deployable cable net may experience unexpected external forces such as wind, snow, temporary load, etc. The closed-loop control enables that structure to adjust the prestress automatically to the target prestress level subject to the external forces, while the open-loop control only deforms passively to equilibrate the external force.

## 6 Conclusion

This paper presents the design and control of a deployable hyperbolic paraboloid cable net. We first give the statics and dynamics equations for the clustered tensegrity structures. Then, the structural topology of the deployable hyperbolic paraboloid cable net actuated by cables is shown. To reduce the number of actuators, we introduce the clustered strategy and apply it to the proposed cable net topology. Based on the deployable clustered cable net design, the deployment trajectory and the actuation prestress in the cables are calculated by the form-finding method. We also check the feasibility and redesign the tensions in the cables during the deployment process. The prestress redesign procedure can be used to avoid cable tensions exceeding their yield strength, eliminate actuating or sensing hardware limitations, or simplify the deployment process for any tensegrity structures.

Moreover, we compute the required rest length as the control variable and propose the open-loop and closed-loop control laws for the cable net. A lab-scale model is constructed to validate the actuation laws. Both numerical and experimental analyses are conducted to verify the control strategies. In the numerical study, we analyzed the pseudo-static and dynamics deployment process. Results show that the longer the control process takes, the closer the dynamics response is to the pseudo-static analysis, which agrees with the physics. In our simulation model, the dynamic time history of member force is generally close to the statics solution if the deployment process is longer than 5 seconds. Finally, we test and compare the open-loop and closed-loop control of the experiment model. Results show that both open-loop and closed-loop control methods can achieve the morphing objectives with sufficient precision. But the open-loop control strategy experiences a much more significant error than the close-loop one. We can also conclude that to maintain the required tensions in the cables during the deployment, and close-loop control law is suggested to be considered in actual applications since the feedback of the tensions can help to keep the stiffness of the structure, reject the disturbance from the environment, and compensate the uncertainty of the experiment model. The design methods proposed in this paper can be used for various deployable tensegrity structures and tensegrity robotics.

## Acknowledgment

The research was supported by the Foundation of Key Laboratory of Space Structures of Zhejiang Province (Grant No. 202102).